\documentclass{article}
\usepackage{spconf}      
\usepackage{amsmath}     
\usepackage{amssymb}     
\usepackage{amsfonts}    
\usepackage{graphicx}    
\usepackage{caption}     
\usepackage{tabularx} 
\usepackage{multirow}

\usepackage{booktabs}

\usepackage{hyperref}

\title{TopSeg: A Multi-Scale Topological Framework for Data-Efficient Heart Sound Segmentation}
\name{Peihong Zhang*,
Zhixin Li*\thanks{Peihong Zhang and Zhixin Li contributed equally.},
Yuxuan Liu, Rui Sang, Yiqiang Cai, Yizhou Tan, Shengchen Li}

\address{School of Advanced Technology, Xi’an Jiaotong-Liverpool University, Suzhou, China}
\ninept
\begin{document}
%

\maketitle
\begin{abstract}

Deep learning approaches for heart-sound segmentation built on time--frequency features can be accurate but often rely on large expert-labeled datasets, limiting robustness and deployment. We present TopSeg, a topological representation-centric framework that encodes Phonocardiogram (PCG) dynamics with multi-scale topological features and decodes them using a lightweight temporal convolutional network (TCN) with an order- and duration-constrained inference step. To evaluate data efficiency and generalization, we train exclusively on PhysioNet 2016 dataset with subject-level subsampling and perform external validation on CirCor dataset. Under matched-capacity decoders, the topological features consistently outperform spectrogram and envelope inputs, with the largest margins at low data budgets; as a full system, TopSeg surpasses representative end-to-end baselines trained on their native inputs under the same budgets while remaining competitive at full data. Ablations at 10\% training confirm that all scales contribute and that combining $H_0$ and $H_1$ yields more reliable S1/S2 localization and boundary stability. These results indicate that topology-aware representations provide a strong inductive bias for data-efficient, cross-dataset PCG segmentation, supporting practical use when labeled data are limited.

\end{abstract}
\section{Introduction}
\label{sec:intro}

\begin{figure}[t]
    \centering
    \includegraphics[width=0.3\textwidth]{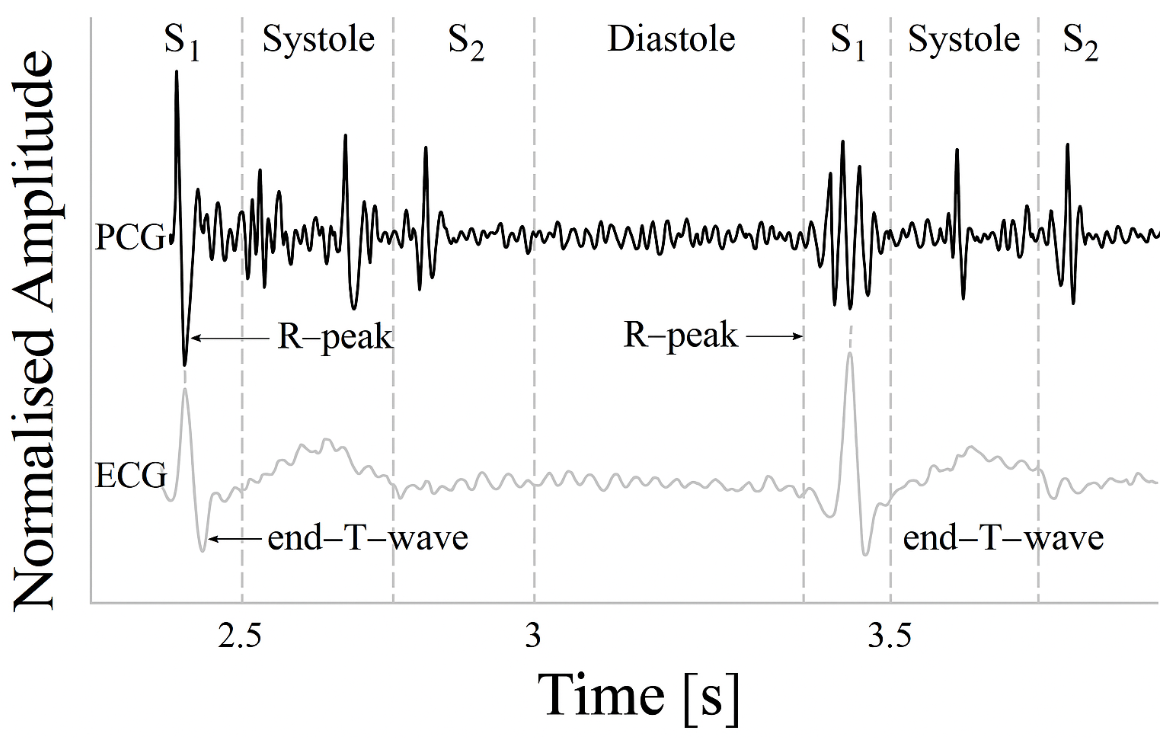}
    \caption{PCG signal segmented into S1, systole, S2, and diastole.}
    \label{fig:pcg_example}

\end{figure}

Heart disease is the leading cause of mortality worldwide, responsible for approximately 17.9 million deaths annually \cite{vos2020global}, with auscultation serving as a primary, cost-effective diagnostic tool in clinical practice \cite{montinari2019first}. Accurate segmentation of phonocardiogram (PCG) signals into their primary components---first (S1) and second (S2) heart sounds, and the intervening systolic and diastolic intervals---aids physicians in cardiac diagnosis by helping identify timing abnormalities, detect murmurs, and assess valvular function \cite{rangayyan1987phonocardiogram,zhang2025nmcse}. Fig.~\ref{fig:pcg_example} illustrates the segmentation task, where PCG signals are segmented into S1, S2, systolic, and diastolic phases.


Recent advances indicate that deep learning models achieve high performance in PCG segmentation by learning representations from spectro-temporal inputs such as spectrograms \cite{he2021research}, mel-spectrograms \cite{fernando2019heart}, and scalograms \cite{gelpud2021deep}, or directly from raw waveforms through end-to-end architectures \cite{chen2021end}. However, when spectro-temporal features are used in deep neural networks, high performance typically relies on large-scale annotated datasets, and performance drops sharply in data-efficient scenarios \cite{zhou2022analysis,Zhang2025}. While handcrafted spectro-temporal features applied to machine learning models can reduce data requirements, they often fail to capture the complex temporal structures of PCGs, thereby limiting their overall effectiveness \cite{chen2021end}.

In addition to spectro-temporal features, some studies have explored envelope-based and statistical representations for PCG segmentation, which require fewer labeled data compared to spectro-temporal approaches \cite{dwivedi2018algorithms}. A widely used approach is the logistic regression hidden semi-Markov model built upon Hilbert and homomorphic envelopes \cite{springer2015logistic}. Variants have since incorporated zero-frequency filtering combined with Hilbert envelopes \cite{prasad2020detection} to improve boundary detection. More recently, segmentation approaches have utilized clustering of envelope-derived features and wavelet envelopes \cite{xu2023optimal} to adaptively locate S1 and S2 events. However, their effectiveness is constrained by inherent limitations: envelope and statistical features are highly sensitive to background noise, prone to misidentification in pathological or irregular heart sounds, and insufficient for capturing the complex temporal–structural patterns present in PCGs \cite{arjoune2024noise}. Consequently, while such features reduce data requirements, their lack of robustness and descriptive power restricts their utility for reliable segmentation in diverse clinical scenarios.


Despite these advances, PCG segmentation remains a challenging task. Annotated datasets are limited because labeling PCG signal requires the expertise of trained clinicians, which makes large-scale annotation difficult \cite{fernando2019heart}. Moreover, the quality of recordings is often compromised by environmental noise and patient-specific factors, further complicating the process of reliable segmentation \cite{chen2021end}. These challenges naturally raise the question: are there alternative feature representations that are robust and fully capture the structural characteristics of PCG signals under data-efficient conditions?

Topological data analysis (TDA) offers a promising direction to address these challenges by extracting features that capture the underlying structural organization of the data. Topological features are inherently robust to small perturbations and noise \cite{carlsson2009topology}, while their topological structural priors reduce reliance on large annotated datasets \cite{leventhal2023exploring}, making them particularly suitable for data-efficient scenarios. This is especially relevant for PCG signals, which are often corrupted by respiratory noise and patient-specific artifacts and available only in limited annotated quantities. Although TDA has been explored in other biomedical domains \cite{skaf2022topological}, its application to PCG segmentation remains unexplored.


In this work, we propose TopSeg, a heart sound segmentation framework that leverages multi-scale topological representations to capture the intrinsic structural patterns of PCGs. Our key insight is that cardiac cycles form persistent structural patterns across multiple physiological timescales, which remain stable under noise perturbations and can be reliably captured even with limited annotated data. To realize this idea, we extract multi-scale topological features at three complementary temporal resolutions: global rhythm patterns (2--8\,s), individual cardiac cycles ($\approx$500\,ms), and fine-grained S1/S2 components ($\approx$100\,ms). These features are incorporated into multiple baseline segmentation models as complementary inputs, enriching their representation capacity. Extensive evaluations on several benchmark datasets demonstrate that integrating multi-scale topological features consistently enhances segmentation accuracy and yields improved generalization under data-efficient conditions.

In summary, our contributions are twofold:
\begin{itemize}
\item We propose TopSeg, a topological framework for PCG segmentation; to our knowledge, this is the first to operationalize multi-scale topological descriptors for this task. These descriptors can be used alone or seamlessly integrated into standard time--frequency baselines with minimal changes.
\item The topological features extracted by TopSeg impart a geometry-aware inductive bias that promotes morphology and scale consistency, underpinning domain generalization and enabling robust cross-dataset transfer under limited data.
\end{itemize}

\section{Topological Data Analysis}

TDA provides multi-scale topological invariants that are provably stable to small perturbations of the input \cite{carlsson2009topology,chazal2016structure}. 
Its central construct, persistent homology (PH), assigns to a filtration $\{\mathcal{K}_\varepsilon\}_{\varepsilon\ge 0}$ of simplicial complexes the collection of birth--death pairs
$D=\{(b_i,d_i)\}$, where $b_i$ (resp.\ $d_i$) denotes the scale at which a homological feature appears (resp.\ disappears). 
Features with large persistence $d_i-b_i$ are typically interpreted as signal; short-lived features are attributed to noise. 
These stability and scale-selection properties make PH attractive for label-efficient signal analysis \cite{leventhal2023exploring}.


To obtain finite-dimensional, differentiable descriptors, we map a persistence diagram $D$ to persistence landscapes \cite{bubenik2015statistical}. 
Each pair $(b,d)\in D$ induces the tent function
\[
f_{(b,d)}(\varepsilon)=\max\!\bigl(0,\,\min(\varepsilon-b,\,d-\varepsilon)\bigr),
\]
and the $k$-th landscape is the pointwise $k$-maximum envelope
\[
\lambda_k(\varepsilon)=\mathrm{kmax}\,\bigl\{\, f_{(b,d)}(\varepsilon) : (b,d)\in D \,\bigr\}.
\]
Sampling $\{\lambda_k\}_{k=1}^{K}$ on a one-dimensional grid of size $G$ yields a fixed-length representation that preserves PH stability in $L^p$ norms and admits straightforward averaging, thereby facilitating integration with learning-based models \cite{bubenik2015statistical}. 


\begin{figure}[t]
    \centering
    \includegraphics[width=0.93\columnwidth]{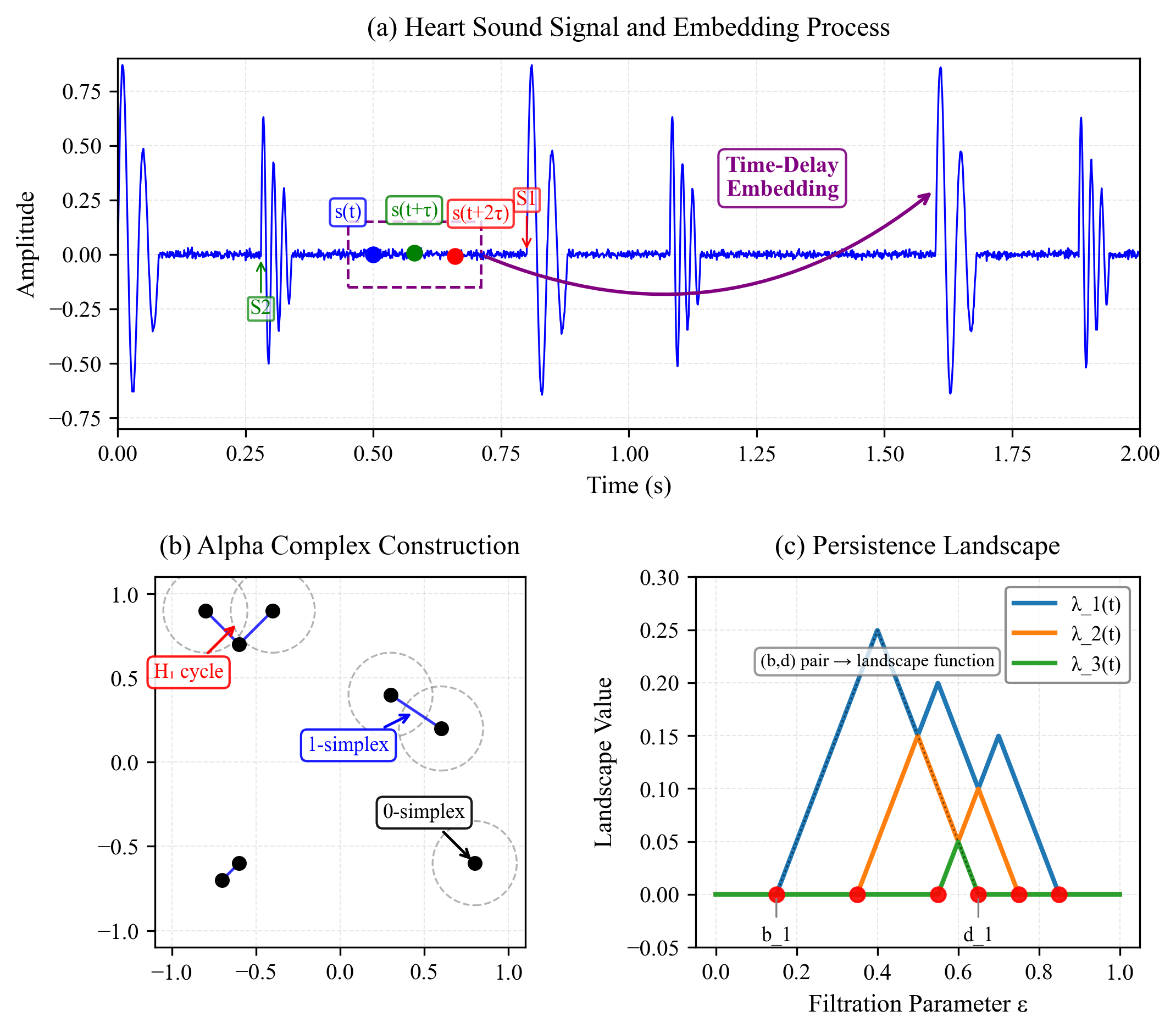}
    \captionsetup{skip=2pt}
    \caption{\textbf{TDA overview.} 
    (a) Time-delay embedding of a heart-sound segment; repeating cardiac cycles induce recurrent structures in the reconstructed phase space. 
    (b) Filtration of simplicial complexes indexed by the scale parameter $\varepsilon$ (illustrated with an $\alpha$-complex), from which PH collects birth--death pairs for $H_0/H_1$. 
    (c) Persistence landscapes $\{\lambda_k\}$ obtained from the diagram provide stable, fixed-length descriptors amenable to learning.}
    \label{fig:tda}
\end{figure}

\section{Proposed TopSeg Framework}

\begin{figure}[t]
\centering
\includegraphics[width=0.41\textwidth]{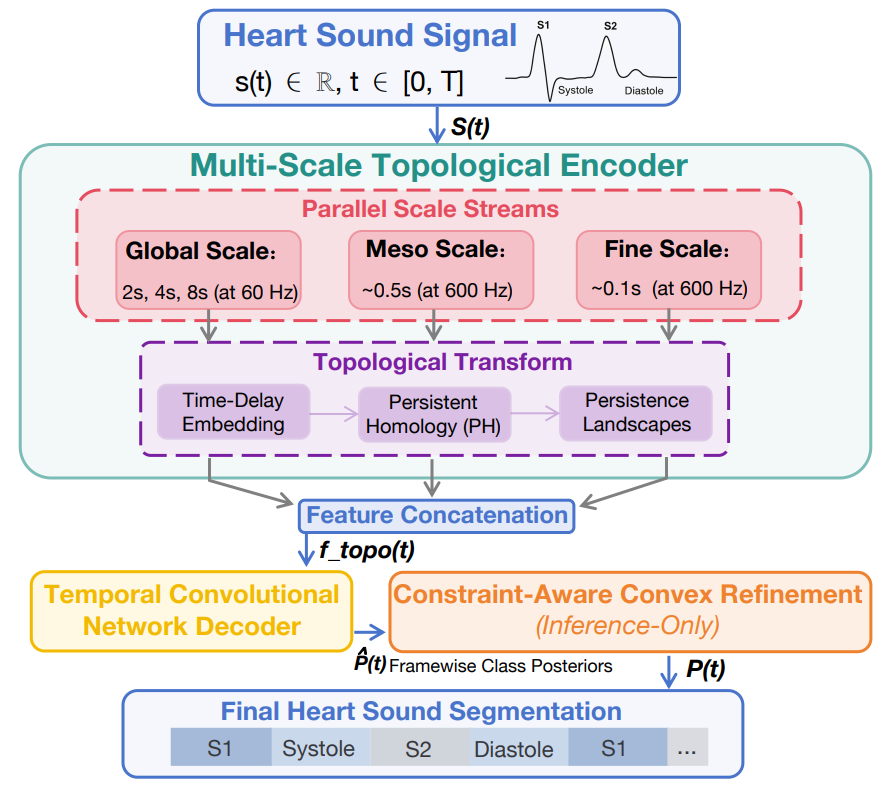}
\caption{
Illustration of the TopSeg framework. 
}

\label{fig:topseg_framework}
\end{figure}

TopSeg has three stages (Fig.~\ref{fig:topseg_framework}): 
(1) multi-scale topological encoder that extracts descriptors via time-delay embedding and persistent homology; 
(2) temporal decoder that maps the topological descriptors to framewise posteriors; and 
(3) inference-time convex refinement that enforces physiological consistency. 

We operationalize TDA’s robustness by using multi-scale persistence descriptors as inputs and coupling them with a lightweight decoder plus a constraint-aware inference layer.

\subsection{Multi-scale Topological Feature Extraction}
\label{sec:topo}

We extract topology-aware descriptors at three physiologically grounded scales: global (multi-beat rhythm), meso (single-cycle morphology), and fine (S1/S2 onsets).

For each scale $\ell\in\{\text{global},\text{meso},\text{fine}\}$, let $s_\ell(t)$ denote the prefiltered waveform. A delay-coordinate embedding with delay $\tau_\ell$ and dimension $d_\ell$ is formed as
\[
\Phi_\ell(t)=\big(s_\ell(t),\,s_\ell(t+\tau_\ell),\,\ldots,\,s_\ell(t+(d_\ell-1)\tau_\ell)\big),
\]
covering an effective window $W_\ell=(d_\ell-1)\tau_\ell$.

\noindent\textbf{Scales and parameterization.}
$W_\ell$ is set by physiological targets. $(\tau_\ell,d_\ell)$ are initialized via average mutual information (AMI) and refined within a small neighborhood to match the intended scale.
To balance context and cost, the global branch operates at $60$\,Hz, whereas meso/fine run at $600$\,Hz, shown in Table~\ref{tab:multiscale_params}.

\begin{table}[t]
\centering
\caption{Multi-scale configuration. $W=(d-1)\tau$. Global descriptors (2/4/8\,s) are averaged into a single global stream.}
\label{tab:multiscale_params}
\footnotesize
\resizebox{\columnwidth}{!}{%
\begin{tabular}{lcccc}
\toprule
\textbf{Scale} & \textbf{$\tau$ (ms)} & \textbf{$d$} & \textbf{$W$ (s)} & \textbf{Physiological target} \\
\midrule
Global-2s (60\,Hz) & 100 & 21 & 2.0 & Multi-beat rhythm \\
Global-4s (60\,Hz) & 200 & 21 & 4.0 & Long context \\
Global-8s (60\,Hz) & 200 & 41 & 8.0 & Rhythm trend \\
Meso (600\,Hz)     & 25  & 21 & 0.50 & One cardiac cycle (60--120\,bpm) \\
Fine (600\,Hz)     & 10  & 11 & 0.10 & S1/S2 morphology (80--150\,ms) \\
\bottomrule
\end{tabular}}
\vspace{-6pt}
\end{table}

\noindent\textbf{From embeddings to topological descriptors.}
Let $\mathcal{P}_\ell=\{\Phi_\ell(t)\}$ be the embedded trajectory. 
A window of length $L_\ell=mW_\ell$ slides with hop equal to one frame (global: $1/60$\,s; meso/fine: $1/600$\,s). 
For each window centered at time $t$, we construct a sparsified Vietoris--Rips filtration from a $k$-NN graph ($k=\lceil\sqrt{n}\rceil$, $n$ points), clipping the filtration radius at the upper distance quantile $q\in[0.90,0.99]$.
Persistent homology is computed in $H_0$ and $H_1$ to obtain birth--death pairs $\mathcal{D}_\ell$, which are mapped to persistence landscapes with $K$ layers sampled on a grid of $G$ points, yielding a fixed-length vector per window.
For the global scale, the 2\,s/4\,s/8\,s landscapes are aggregated by element-wise mean into a single global descriptor.
Window-centered descriptors are assigned to the center frame; linear interpolation across overlaps mitigates boundary drift and yields smoothly varying per-frame features.
Concatenating the global, meso, and fine descriptors gives
\[
\mathbf{f}_{\text{topo}}(t)\in\mathbb{R}^{D_{\text{topo}}},\qquad 
D_{\text{topo}}=3\times(2KG),
\]
with each scale contributing $2KG$ dimensions (both $H_0$ and $H_1$).

\subsection{Topology-Only Constraint-Aware Refinement}
\label{sec:fusion_refine}

We refine the framewise class posteriors produced by the topological decoder via a small convex program that encodes physiological priors. Training optimizes the network on raw posteriors; refinement is applied only at inference.

\noindent\textbf{Topology target.}
From fine-scale landscapes $\{\lambda_{\text{fine},h}^{(k)}(t)\}$ ($h\!\in\!\{0,1\}$) we build
\[
r_{\mathrm{topo}}(t)=\mathrm{Norm}\!\left(\sum_{k}\lambda_{\text{fine},0}^{(k)}(t)+\sum_{k}\lambda_{\text{fine},1}^{(k)}(t)\right)\in[0,1],
\]
where $\mathrm{Norm}(\cdot)$ is percentile-based within a short sliding window. 
$r_{\mathrm{topo}}(t)$ serves as a fixed, non-trainable guide.

\noindent\textbf{Topology-derived reliability.}
To modulate the alignment strength using only topology, we define
\[
\eta(t)=\sigma\!\big(\gamma\,[\,r_{\mathrm{topo}}^{\mathrm{EMA}}(t)-\tau\,]\big)\in[0,1],
\]
where $r_{\mathrm{topo}}^{\mathrm{EMA}}(t)=\rho\,r_{\mathrm{topo}}^{\mathrm{EMA}}(t{-}1)+(1{-}\rho)\,r_{\mathrm{topo}}(t)$ is an EMA with decay $\rho$ (initialized at $r_{\mathrm{topo}}(0)$), $\tau$ is a mid-level threshold, and $\gamma$ controls the sharpness.

\noindent\textbf{Refinement objective (inference-only).}
Let $P_{\mathrm{evt}}(t)=P_{\mathrm{S1}}(t)+P_{\mathrm{S2}}(t)$. For each recording we refine $\hat{P}$ by solving
\begin{equation}
\label{eq:refine}
\begin{aligned}
\min_{\{P(t)\}}\quad 
& \sum_{t} \|P(t)-\hat{P}(t)\|_2^2 
  + \lambda_s \sum_{t} \|P(t)-P(t{-}1)\|_2^2  \\
& + \lambda_b \sum_{t} \big(P_{\mathrm{evt}}(t)-\theta_{\max}\big)_+^2 \\
& + \lambda \sum_{t} \eta(t)\, \|P_{\mathrm{evt}}(t)-r_{\mathrm{topo}}(t)\|_2^2 , \\
\text{s.t.}\quad 
& P(t)\in\Delta^{C}\ (C{=}4),\ \forall t ,
\end{aligned}
\end{equation}
with $(\cdot)_+=\max(0,\cdot)$. 
The terms respectively (i) keep $P$ close to the network posterior, (ii) impose Tikhonov temporal smoothness, (iii) cap instantaneous event mass, and (iv) softly align $P_{\mathrm{evt}}$ with topology peaks under the topology-derived reliability $\eta(t)$. 
All penalties are convex in $P$; with $\lambda_s>0$ and simplex projection, the objective is strongly convex and admits a unique minimizer. 
We use projected proximal gradient; temporal coupling is tridiagonal, so each iteration costs $O(TC)$. A small fixed number of iterations is used per recording (Tab.~\ref{tab:topseg_hparams}).

\begin{table}[t]
\scriptsize
\setlength{\tabcolsep}{3.2pt}
\renewcommand{\arraystretch}{1.05}
\centering
\caption{TopSeg hyperparameters (defaults and compact ranges).}
\label{tab:topseg_hparams}
\begin{tabularx}{\columnwidth}{l c c X}
\toprule
\textbf{Block} & \textbf{Symbol} & \textbf{Default} & \textbf{Notes / Range} \\
\midrule
Embedding & $W_\ell$ & Tab.~\ref{tab:multiscale_params} & physiology \\
(per scale) & $L_\ell$ & $mW_\ell$ & $m{=}2$ (1.5--2.5) \\
\midrule
PH pipeline & $k$ & $\lceil\sqrt{n}\rceil$ & k-NN \\
(per window) & $q$ & 0.95 & 0.90--0.99 \\
& $(K,G)$ & (5,128) & landscape layers \\
\midrule
Topology cues & $\eta(t)$ & $\sigma(\gamma[r_{\mathrm{topo}}^{\mathrm{EMA}}{-}\tau])$ & $\gamma{=}2.0,\ \tau{=}0.5,\ \rho{=}0.90$ \\
\midrule
Refine \eqref{eq:refine} & $\lambda_s$ & $1{\times}10^{-2}$ & $[5{\times}10^{-3},2{\times}10^{-2}]$ \\
& $\lambda_b$ & $5{\times}10^{-2}$ & $[2{\times}10^{-2},8{\times}10^{-2}]$ \\
& $\lambda$ & $5{\times}10^{-2}$ & $[2{\times}10^{-2},8{\times}10^{-2}]$ \\
& $\theta_{\max}$ & 0.65 & event-mass cap \\
\midrule
Solver & $N_{\!iter}$ & 8 & monotone PGD \\
\bottomrule
\end{tabularx}
\vspace{-4pt}
\end{table}




\section{Experiment and Result}


\subsection{Dataset}

We train on the PhysioNet/CinC 2016 database \cite{liu2016open} and simulate data-efficient learning by randomly subsampling subjects at 5\%, 10\%, 25\%, 50\%, and 100\% of the training pool; the corpus comprises 3,153 recordings from 764 subjects. For external validation, we use the CirCor DigiScope Dataset \cite{oliveira2022circor} with 5,272 recordings from 1,568 subjects. Since PhysioNet 2016 does not provide framewise segmentation for all files, we obtain S1/S2 labels by using the official logistic-regression HSMM segmenter released with the Challenge \cite{springer2015logistic} and, where available, we preferentially adopt the hand-corrected training annotations.

\begin{table*}[t]
\centering
\scriptsize
\setlength{\tabcolsep}{5pt}
\begin{tabular}{lccccc}
\toprule
\multicolumn{6}{c}{\textbf{CirCor DigiScope external validation} (macro-F1, \%, 60\,ms tolerance) }\\
\cmidrule(lr){1-6}
Model & Train 5\% & Train 10\% & Train 25\% & Train 50\% & Train 100\% \\
\midrule
\multicolumn{6}{l}{\textit{Q1: Feature-controlled (same decoder; different inputs)}}\\
MLP (envelope)                                   & 43.0 & 48.1 & 54.3 & 58.2 & 62.5 \\
MLP (log-mel)                                    & 48.4 & 53.2 & 60.5 & 64.7 & 67.9 \\
\textbf{MLP (topological, ours)}                 & \textbf{54.6} & \textbf{59.3} & \textbf{65.2} & \textbf{69.1} & \textbf{72.8} \\
TCN (envelope)            & 55.2 & 61.3 & 67.7 & 71.6 & 75.4 \\
TCN (log-mel)             & 58.1 & 64.2 & 71.3 & 76.4 & 80.2 \\
\textbf{TCN (topological, ours)} & \textbf{64.1} & \textbf{70.4} & \textbf{76.2} & \textbf{80.5} & \textbf{83.1} \\
\midrule
\multicolumn{6}{l}{\textit{Q2: End-to-end comparators (native inputs; same budgets)}}\\
LR--HSMM (envelope, native) \cite{springer2015logistic}   & 54.3 & 57.2 & 60.3 & 63.2 & 66.1 \\
CLSTM (raw audio, native) \cite{chen2021end}              & 58.4 & 64.5 & 72.3 & 77.2 & 81.3 \\
U\,-\,Net (time--frequency, native) \cite{he2021research} & 59.2 & 65.3 & 73.1 & 78.4 & 82.5 \\
FFT+CNN U\,-\,Net (time--frequency) \cite{park2025enhancement} & 60.5 & 66.3 & 74.2 & 79.4 & 83.2 \\
\textbf{Ours: Topological features + TCN (order/duration-constrained decoding)} & \textbf{66.7} & \textbf{71.9} & \textbf{78.4} & \textbf{82.1} & \textbf{85.3} \\
\bottomrule
\end{tabular}
\caption{External validation on CirCor with models trained on PhysioNet under different data budgets.}
\label{tab:circor_external_only}
\end{table*}

\subsection{Data Preprocessing}
PCG recordings are first band-limited to 20--200\,Hz using a zero-phase Butterworth band-pass filter \cite{debbal2008computerized}, and then downsampled to 600\,Hz with a polyphase anti-aliasing decimator to prevent aliasing. Signals are z-score normalized within each recording. For training we use 10\,s segments: longer recordings are split into non-overlapping 10\,s chunks, while shorter ones are looped to 10\,s to preserve batch shape without changing local dynamics. The 20--200\,Hz passband preserves the dominant energy of S1/S2 while suppressing low-frequency motion artifacts and high-frequency ambient noise.

\subsection{Comparison Models}
To demonstrate both the data-efficiency of our representation and the overall effectiveness of our pipeline, we compare along two complementary axes. First, in a feature-controlled setting, we train the same Temporal Convolutional Network (TCN) \cite{lea2017temporal} decoder and a shallow MLP (1 hidden layer, 128 units) on three per-frame inputs—our topological landscapes, log-mel spectrograms (64 bins), and Hilbert envelope—thereby isolating the contribution of representations under identical model capacity. Second, at the architecture level, we include end-to-end models trained on their native inputs: the envelope-based LR–HSMM \cite{springer2015logistic}, a time–frequency CLSTM \cite{chen2021end}, a U-Net–style TF model \cite{he2021research}, and an FFT+CNN U-Net \cite{park2025enhancement}. All models use the same optimizer, subject-level splits, early stopping, and a 60\,ms boundary tolerance; when indicated, we apply the same convex refinement at inference to enforce left-to-right ordering and minimal durations. This protocol tests (i) whether our representation yields superior accuracy in low-data regimes when decoded by matched-capacity models, and (ii) whether our full system outperforms representative architectures under identical data budgets.





\subsection{Results under Data-Efficient Training}
Table~\ref{tab:circor_external_only} summarizes external validation on CirCor (four-state macro-F1 with a 60\,ms boundary tolerance), with all models trained on PhysioNet under subject-level subsampling and identical optimization. 

\textbf{Q1 (Representation).} Under matched-capacity decoders (shallow MLP and the same TCN), our topological features dominate log-mel and envelope across all data budgets. The advantage is most pronounced in the low-data regime (5–25\%) and remains positive at full data, indicating that the representation carries stronger inductive bias and learns reliably with limited supervision.

\textbf{Q2 (Overall model).} Our full pipeline—topological features decoded by a lightweight TCN with order/duration-constrained decoding—outperforms representative end-to-end baselines trained on their native inputs at every budget, with the largest margins when data are scarce and competitive performance at 100\%. This pattern holds under cross-dataset evaluation (adult/mixed PhysioNet $\rightarrow$ pediatric, multi-location CirCor), supporting the claim that our approach is both data-efficient and robust to domain shift. As expected, absolute scores are below in-domain reports on CirCor, reflecting the stricter generalization setting rather than a weakness of the method.

\subsection{Multi-Scale Ablation}
We quantify the contribution of each temporal scale in our topological representation. 
Starting from the full three-branch design (global, meso, fine), we remove one branch at a time while keeping the rest of the pipeline unchanged. 
All ablations use the same decoder (the TCN in \cite{lea2017temporal} with identical capacity and training schedule) and the same order/duration-constrained decoding at inference for fairness. 
Models are trained on 10\% of PhysioNet/CinC 2016 (subject-level subsampling; three random draws) and evaluated on CirCor with four-state macro-F1 (\%) under a 60\,ms boundary tolerance.

\begin{table}[t]
\centering
\scriptsize
\setlength{\tabcolsep}{4pt}
\begin{tabular}{lccc}
\toprule
\multicolumn{4}{c}{\textbf{Ablation at 10\% training } \;\; macro-F1 }\\
\cmidrule(lr){1-4}
Variant & Overall & S1 & S2 \\
\midrule
Full: global + meso + fine   & \textbf{71.9} & \textbf{79.0} & \textbf{77.6} \\
\quad w/o global             & 70.9 & 78.2 & 76.9 \\
\quad w/o meso               & 70.5 & 77.1 & 75.9 \\
\quad w/o fine               & 69.8 & 74.4 & 73.2 \\
$H_0$-only (no $H_1$)        & 68.3 & 75.1 & 73.7 \\
$H_1$-only (no $H_0$)        & 66.8 & 73.4 & 72.1 \\
\bottomrule
\end{tabular}
\caption{Multi-scale and homology ablations under the 10\% budget.}
\label{tab:ablation_scales}
\end{table}

\noindent\textbf{Analysis.}
The ablation in Table~\ref{tab:ablation_scales} clarifies why our representation is data-efficient.
(1) \emph{Each scale is necessary}: removing any branch consistently degrades macro-F1, confirming that global (slow rhythm), meso (single-cycle structure), and fine (sharp transients) capture complementary cues.
(2) \emph{Fine $\rightarrow$ onsets}: dropping the fine branch produces the largest losses in S1/S2, indicating its role in precise valve-closure localization under noise.
(3) \emph{Meso $\rightarrow$ intra-cycle}: removing the meso branch mainly harms systole/diastole delineation, consistent with its coverage of one–few cardiac cycles.
(4) \emph{Global $\rightarrow$ stability}: the global branch improves long-range temporal consistency and reduces short-state oscillations; its removal yields smaller but systematic declines.
(5) \emph{$H_0$ and $H_1$ are complementary}: either homology alone underperforms the combined design, showing that both amplitude-driven morphology ($H_0$) and loop-level patterns ($H_1$) are informative.


\noindent\textbf{Practicality.}
For deployment, topological features are computed once per recording and cached; test-time decoding uses the same lightweight TCN and the convex refinement from Sec.~3.2, adding negligible latency.
Runtime scales near-linearly with sequence length and graph sparsity, and admits a smooth speed–accuracy trade-off by reducing landscape sampling density.

\section{Conclusion}
We introduced TopSeg, a data-efficient PCG segmentation framework that couples multi-scale topological descriptors with a lightweight TCN and a constraint-aware inference step. In cross-dataset evaluations with device and patient shifts, TopSeg consistently surpasses feature- and architecture-level baselines, especially with limited labels. Ablations trace these gains to complementary global/meso/fine scales and $H_0/H_1$, which confer a geometry-aware inductive bias enabling domain generalization and robust transfer—making TopSeg practical and easy to integrate for real-world deployment under label scarcity and device variability.

\newpage

\section{Acknowledgment}
This work was funded by Basic Research Program of Jiangsu Province under Grant BG2024027.

\bibliographystyle{IEEEbib}
\bibliography{refs}

@article{vos2020global,
  title={Global burden of 369 diseases and injuries in 204 countries and territories, 1990--2019: a systematic analysis for the Global Burden of Disease Study 2019},
  author={Vos, Theo and Lim, Stephen S and Abbafati, Cristiana and Abbas, Kaja M and Abbasi, Mohammad and Abbasifard, Mitra and Abbasi-Kangevari, Mohsen and Abbastabar, Hedayat and Abd-Allah, Foad and Abdelalim, Ahmed and others},
  journal={The lancet},
  volume={396},
  number={10258},
  pages={1204--1222},
  year={2020},
  publisher={Elsevier}
}

@article{montinari2019first,
  title={The first 200 years of cardiac auscultation and future perspectives},
  author={Montinari, Maria Rosa and Minelli, Sergio},
  journal={Journal of multidisciplinary healthcare},
  pages={183--189},
  year={2019},
  publisher={Taylor \& Francis}
}

@article{rangayyan1987phonocardiogram,
  title={Phonocardiogram signal analysis: a review.},
  author={Rangayyan, Rangraj M and Lehner, Richard J},
  journal={Critical reviews in biomedical engineering},
  volume={15},
  number={3},
  pages={211--236},
  year={1987}
}

@article{springer2015logistic,
  title={Logistic regression-HSMM-based heart sound segmentation},
  author={Springer, David B and Tarassenko, Lionel and Clifford, Gari D},
  journal={IEEE transactions on biomedical engineering},
  volume={63},
  number={4},
  pages={822--832},
  year={2015},
  publisher={IEEE}
}

@article{fernando2019heart,
  title={Heart sound segmentation using bidirectional LSTMs with attention},
  author={Fernando, Tharindu and Ghaemmaghami, Houman and Denman, Simon and Sridharan, Sridha and Hussain, Nayyar and Fookes, Clinton},
  journal={IEEE journal of biomedical and health informatics},
  volume={24},
  number={6},
  pages={1601--1609},
  year={2019},
  publisher={IEEE}
}

@article{liu2016open,
  title={An open access database for the evaluation of heart sound algorithms},
  author={Liu, Chengyu and Springer, David and Li, Qiao and Moody, Benjamin and Juan, Ricardo Abad and Chorro, Francisco J and Castells, Francisco and Roig, Jos{\'e} Millet and Silva, Ikaro and Johnson, Alistair EW and others},
  journal={Physiological measurement},
  volume={37},
  number={12},
  pages={2181},
  year={2016},
  publisher={IOP Publishing}
}

@article{chen2021end,
  title={End-to-end heart sound segmentation using deep convolutional recurrent network},
  author={Chen, Yao and Sun, Yanan and Lv, Jiancheng and Jia, Bijue and Huang, Xiaoming},
  journal={Complex \& Intelligent Systems},
  volume={7},
  pages={2103--2117},
  year={2021},
  publisher={Springer}
}

@article{he2021research,
  title={Research on segmentation and classification of heart sound signals based on deep learning},
  author={He, Yi and Li, Wuyou and Zhang, Wangqi and Zhang, Sheng and Pi, Xitian and Liu, Hongying},
  journal={Applied Sciences},
  volume={11},
  number={2},
  pages={651},
  year={2021},
  publisher={MDPI}
}

@article{carlsson2009topology,
  title={Topology and data},
  author={Carlsson, Gunnar},
  journal={Bulletin of the American Mathematical Society},
  volume={46},
  number={2},
  pages={255--308},
  year={2009}
}

@book{chazal2016structure,
  title={The structure and stability of persistence modules},
  author={Chazal, Fr{\'e}d{\'e}ric and De Silva, Vin and Glisse, Marc and Oudot, Steve},
  volume={10},
  year={2016},
  publisher={Springer}
}

@article{bubenik2015statistical,
  title={Statistical topological data analysis using persistence landscapes.},
  author={Bubenik, Peter and others},
  journal={J. Mach. Learn. Res.},
  volume={16},
  number={1},
  pages={77--102},
  year={2015}
}

@inproceedings{gelpud2021deep,
  title={Deep learning for heart sounds classification using scalograms and automatic segmentation of PCG signals},
  author={Gelpud, John and Castillo, Silvia and Jojoa, Mario and Garcia-Zapirain, Begonya and Achicanoy, Wilson and Rodrigo, David},
  booktitle={International Work-Conference on Artificial Neural Networks},
  pages={583--596},
  year={2021},
  organization={Springer}
}

@article{zhou2022analysis,
  title={On the analysis of data augmentation methods for spectral imaged based heart sound classification using convolutional neural networks},
  author={Zhou, George and Chen, Yunchan and Chien, Candace},
  journal={BMC medical informatics and decision making},
  volume={22},
  number={1},
  pages={226},
  year={2022},
  publisher={Springer}
}

@inproceedings{prasad2020detection,
  title={Detection of S1 and S2 locations in phonocardiogram signals using zero frequency filter},
  author={Prasad, RaviShankar and Yilmaz, Gurkan and Chetelat, Olivier and Doss, Mathew Magimai-},
  booktitle={ICASSP 2020-2020 IEEE International Conference on Acoustics, Speech and Signal Processing (ICASSP)},
  pages={1254--1258},
  year={2020},
  organization={IEEE}
}

@article{xu2023optimal,
  title={Optimal heart sound segmentation algorithm based on k-mean clustering and wavelet transform},
  author={Xu, Xingchen and Geng, Xingguang and Gao, Zhixing and Yang, Hao and Dai, Zhiwei and Zhang, Haiying},
  journal={Applied Sciences},
  volume={13},
  number={2},
  pages={1170},
  year={2023},
  publisher={MDPI}
}

@article{arjoune2024noise,
  title={A noise-robust heart sound segmentation algorithm based on Shannon energy},
  author={Arjoune, Youness and Nguyen, Trong N and Doroshow, Robin W and Shekhar, Raj},
  journal={IEEE Access},
  volume={12},
  pages={7747--7761},
  year={2024},
  publisher={IEEE}
}

@article{dwivedi2018algorithms,
  title={Algorithms for automatic analysis and classification of heart sounds--a systematic review},
  author={Dwivedi, Amit Krishna and Imtiaz, Syed Anas and Rodriguez-Villegas, Esther},
  journal={IEEE Access},
  volume={7},
  pages={8316--8345},
  year={2018},
  publisher={IEEE}
}

@article{leventhal2023exploring,
  title={Exploring classification of topological priors with machine learning for feature extraction},
  author={Leventhal, Samuel and Gyulassy, Attila and Heimann, Mark and Pascucci, Valerio},
  journal={IEEE Transactions on Visualization and Computer Graphics},
  volume={30},
  number={7},
  pages={3959--3972},
  year={2023},
  publisher={IEEE}
}

@article{skaf2022topological,
  title={Topological data analysis in biomedicine: A review},
  author={Skaf, Yara and Laubenbacher, Reinhard},
  journal={Journal of Biomedical Informatics},
  volume={130},
  pages={104082},
  year={2022},
  publisher={Elsevier}
}

@inproceedings{oliveira2022circor,
  title={The circor digiscope phonocardiogram dataset},
  author={Oliveira, Jorge and Renna, Francesco and Costa, Paulo and Nogueira, Marcelo and Oliveira, Ana Cristina and Elola, Andoni and Ferreira, Carlos and Jorge, Alipio and Rad, Ali Bahrami and Reyna, Matthew and others},
  booktitle={IEEE Conference},
  year={2022}
}

@article{debbal2008computerized,
  title={Computerized heart sounds analysis},
  author={Debbal, SM and Bereksi-Reguig, Fethi},
  journal={Computers in biology and medicine},
  volume={38},
  number={2},
  pages={263--280},
  year={2008},
  publisher={Elsevier}
}

@article{park2025enhancement,
  title={Enhancement of phonocardiogram segmentation using convolutional neural networks with Fourier transform module},
  author={Park, Changhyun and Shin, Keewon and Seo, Jinew and Lim, Hyunseok and Kim, Gyeong Hoon and Seo, Woo-Young and Kim, Sung-Hoon and Kim, Namkug},
  journal={Biomedical Engineering Letters},
  volume={15},
  number={2},
  pages={401--413},
  year={2025},
  publisher={Springer}
}

@inproceedings{lea2017temporal,
  title={Temporal convolutional networks for action segmentation and detection},
  author={Lea, Colin and Flynn, Michael D and Vidal, Rene and Reiter, Austin and Hager, Gregory D},
  booktitle={proceedings of the IEEE Conference on Computer Vision and Pattern Recognition},
  pages={156--165},
  year={2017}
}

@inproceedings{Zhang2025,
    author = "Zhang, Peihong and Liu, Yuxuan and Li, Zhixin and Sang, Rui and Tan, Yizhou and Cai, Yiqiang and Li, Shengchen",
    title = "An Entropy-Guided Curriculum Learning Strategy for Data-Efficient Acoustic Scene Classification under Domain Shift",
    booktitle = "Proceedings of the 10th Workshop on Detection and Classification of Acoustic Scenes and Events (DCASE 2025)",
    address = "Barcelona, Spain",
    month = "October",
    year = "2025",
    pages = "100--104",
    doi = "10.5281/zenodo.17251589"
}

@article{zhang2025nmcse,
  title={NMCSE: Noise-robust multi-modal coupling signal estimation method via optimal transport for cardiovascular disease detection},
  author={Zhang, Peihong and Li, Zhixin and Sang, Rui and Liu, Yuxuan and Cai, Yiqiang and Tan, Yizhou and Li, Shengchen},
  journal={arXiv preprint arXiv:2505.18174},
  year={2025}
}

\end{document}